\documentclass{revtex4}
\usepackage{latexsym}
\usepackage{amsfonts}
\usepackage[latin1]{inputenc}
\usepackage[caption=false]{subfig}
\usepackage{graphicx}
\usepackage{mathrsfs}
\usepackage{amssymb}
\usepackage{amsmath}
\usepackage{verbatim}
\usepackage{multirow}
\usepackage{color}
\usepackage{epsfig}
\usepackage{amscd}
\usepackage{bm}
\usepackage{fancyhdr}
\usepackage{color}

\begin{document}

\title{A 4D Composite Higgs Model: Testing its Scalar Sector at the LHC}

%

\author{S. Moretti$^\dagger$, A. Belyaev, M.S. Brown and D. Barducci}
\affiliation{School of Physics \& Astronomy, University of Southampton, Southampton SO17 1BJ, UK}
\author{S. De Curtis}
\affiliation{INFN, Sezione di Firenze, Via G. Sansone 1, 50019 Sesto Fiorentino, ITALY}
\author{G.M. Pruna}
\affiliation{TU Dresden, Institut f\"ur Kern- und Teilchenphysik, Zellescher Weg 19, D-01069 Dresden, GERMANY}

\begin{abstract}
We explain the current Large Hadron Collider (LHC)
data pointing to the discovery of a neutral Higgs boson in the context 
of a 4-Dimensional Composite Higgs Model (4DCHM). The full particle spectrum 
of this scenario is derived without any approximation and implemented 
in automated computational tools to enable fast phenomenological investigation.
Several parameter configurations compliant with experimental 
constraints are presented and discussed. A $\chi^2$ fit to the LHC data 
quantifying the consistency of the 4DCHM as a whole with experimental 
evidence is finally performed.
\end{abstract}

\maketitle

\thispagestyle{fancy}


\section{Introduction}
The first question to ask following the discovey of  a Higgs-like signal at the LHC by the 
ATLAS and CMS experiments \cite{:2012gk,:2012gu}\footnote{Some supplemental evidence also emerged at the Tevatron \cite{Aaltonen:2012qt}.} would be whether such an object is fundamental or composite. After all, all (pseudo)scalar
particles so far discovered in nature have been bound states of fermions. On the one hand, plenty of
literature has ignored asking such a question and simply plunged into exploring the innumerable (and unquotable here)
variations of the fundamental Higgs hypothesis. On the other hand, all those who did it eventually considered 
composite Higgs scenarios in some easily calculable regime, whereby the additional particle spectrum
entering the ensuing models (generally comprising both new heavy gauge bosons and fermions) is essentially decoupled,
i.e., by essentially accounting for the new heavy states in their infinite mass limit and studying the residual effects onto the SM sector. This may not be sufficiently accurate if one notices the following. Firstly, both species of new particles can affect the mixing pattern of the Higgs boson, modifying its couplings to the SM particles in a way that may depend on the new gauge boson and 
fermion masses. Secondly,  they can appear as virtual objects interacting with the Higgs boson active at the LHC, if one realises that the Higgs production channel to which the LHC is most sensitive for a mass around 125 GeV is gluon-gluon fusion (which can occur in such models via not only loops of $t$, $b$ quarks but also via $t'$, $b'$ ones) and that the decay channel that appears most anomalous is the photon-photon one (which can occur in such models via not only loops of $t$, $b$ quarks and $W$ bosons but also via $t'$, $b'$ and $W'$ ones). Here too, if the masses of the new objects is not much larger than the Higgs mass, one should expect a dependence on these in the loop functions. 
 In essence, it is clear that a more rigorous approach may be needed.

We followed this approach in Ref.~\cite{Barducci:2013wjc} and we briefly report on it here.
 We will prove that the exact results can deviate from those obtained in the aforementioned decoupling limit of the
new gauge and matter states,
by adopting a particular composite Higgs model for which we have derived exactly the spectrum of particle masses and couplings which intervene at the LHC. Then, by exploiting the latter, we will explore the
viability of the composite Higgs boson hypothesis at the LHC by comparing the corresponding preditions for cross sections and Branching Ratios (BRs) against the ATLAS and CMS data (herein we will neglect Tevatron results). 

This write up is organised as follows. In Sect.~\ref{Sec:Model}, we introduce the reference model adopted and 
briefly describe its Higgs sector. In Sect.~\ref{Sec:strategy} we map its allowed parameter space in the light of the latest experimental results. In Sect.~\ref{Sec:Results} we present our main findings. Finally, we conclude in Sect.~\ref{Sec:Conclusions}.


\section{The Higgs sector of the 4DCHM}
\label{Sec:Model}
\noindent

The model we adopted for our analysis is the 4DCHM of \cite{DeCurtis:2011yx}, to which we refer for further details. 
Our main interest here will be  the composite Higgs state 
and its couplings to both the SM particles (mainly to the $W$ and $Z$ gauge bosons plus the $t$ and $b$ quarks) and to the other new objects belonging to such model (the $W'$ and $Z'$ gauge bosons plus  the $t'$ and $b'$ quarks).
For a detailed phenomenological analysis of the gauge sector of the 4DCHM we refer to \cite{Barducci:2012kk,Barducci:2012as,Barducci:2012sk}.

The 4DCHM is an effective low-energy Lagrangian approximation that represents an extremely deconstructed version of the Minimal Composite Higgs model (MCHM) of \cite{Agashe:2004rs} based on the coset $SO(5)/SO(4)$ that gives four 
Pseudo Nambu-Goldstone Bosons (PNGBs) in the vector representation of $SO(4)$, one of which is the 
(physical) composite Higgs boson, $H$. This extreme deconstruction of the 5D theory leads to a two site schematic representation, respectively called elementary and composite sectors (considered already in  \cite{Contino:2006nn} where, however, the full gauge/Goldstone boson structure of the theory is not incorporated). Although extreme, this two site truncation represents the framework where to study in a computable  way the lowest lying resonances (both bosonic and fermionic) that are the only ones that may be accessible at the LHC. In essence, the 4DCHM represents the ideal phenomenological framework where to test the idea of a composite Higgs boson as a PNGB (see also \cite{Panico:2011pw} although with a different construction).

The composite Higgs particle acquires mass, $m_H$, through a one-loop generated
potential (\`a la Coleman-Weinberg). 
 The particular choice for the fermionic sector of \cite{DeCurtis:2011yx} gives a finite potential and from the location of the minimum one extracts the expression for  $m_H$ and its Vacuum Expectation Value 
 (VEV), $\langle h \rangle$, in terms of the parameters of the model. Further, for a natural choice of these, the Higgs mass can be consistent with the recent results of the  ATLAS \cite{:2012gk} and CMS \cite{:2012gu} experiments, measuring $m_H$ around $125$ GeV (as mentioned). Also for this reason then we will adopt the effective description of the 4DCHM for our phenomenological analysis of Higgs processes at the LHC. Finally,
in the spirit of partial compositeness, spin 1 and spin 1/2 particles from the SM are coupled to the Higgs boson only via mixing with the corresponding composite particles while the new gauge and fermionic resonances directly interact with the Higgs 
field\footnote{In the 4DCHM, regarding the fermionic sector, only the top and bottom quarks are mixed with the composite fermions.}.


\section{The 4DCHM parameter space}
\label{Sec:strategy}
\noindent

Alonside the customary SM matter and force states (the $e^-$, $\mu^-$, $\tau^-$, $\nu_{e,\mu,\tau}$ leptons, the $u$, $d$, $c$, $s$, $t$, $b$ quarks, the $\gamma$, $Z$, $W^{\pm}$, $g$ gauge bosons), the 4DCHM incorporates the Higgs state $H$ as a PNGB and a large number of new particles, both in the fermionic (quark) and bosonic (gauge)  sector: see
Tab.~\ref{table:partspec}.

\begin{table}[t!]
\begin{center}
\begin{tabular}{|l|l|}
\hline
Neutral Gauge Bosons & $Z_{1,2,...,5}$\\
Charged Gauge Bosons & $W_{1,2,3}^{\pm}$\\
Charge 2/3 quarks    & $T_{1,2,...,8}$ \\
Charge $-1/3$ quarks    & $B_{1,2,...,8}$ \\
Charge 5/3 quarks    & $\tilde{T}_{1,2}$ \\
Charge $-4/3$ quarks    & $\tilde{B}_{1,2}$ \\
\hline
\end{tabular}
\end{center}
\caption{Extra particles of the 4DCHM with respect to the SM (an increasing particle number implies a larger mass).}
\label{table:partspec}
\end{table}

To enable an efficient phenomenological analysis of the Higgs sector, we have implemented the 4DCHM in LanHEP \cite{Semenov:2010qt}, through the SLHA+ library \cite{Belanger:2010st}, thereby deriving in an automated way the Feynman rules  in CalcHEP format \cite{Pukhov:1999gg,Belyaev:2012qa}. In addition, we have listed in Tabs.~1 and 2 of Ref.~\cite{Barducci:2012kk}  the correspondence between the model {notations} used  here and in \cite{Barducci:2013wjc} and the ones uploaded onto the High Energy Physics Model Data-Base (HEPMDB) \cite{Brooijmans:2012yi} at  {\tt http://hepmdb.soton.ac.uk/hepmdb:0213.0123} under the name ``4DCHM (with HAA/HGG)''.

To map the valid parameter space of the 4DCHM we have written a Mathematica routine \cite{mathematica}, which considers $f$ (the compositeness scale)
and $g_*$ (the coupling common to the non-SM gauge groups) as free parameters, performs scans over $m_*$, $\Delta_{tL}$, $\Delta_{tR}$, $Y_T$, $M_{Y_T}$, $\Delta_{bL}$, $\Delta_{bR}$, $Y_B$, $M_{Y_B}$ 
(see  \cite{Barducci:2013wjc} for their definition) and finds allowed points reproducing the following 
physical observables: $e$, $M_Z$, $G_F$, with values as per Particle Data Group (PDG)  listing \cite{PDG}, 
$165 ~{\rm GeV} \le m_t \le 175 ~{\rm GeV}$,  $2 ~{\rm GeV} \le m_b \le 6 ~{\rm GeV}$ and  $124 ~{\rm GeV} \le m_H \le 126 ~{\rm GeV}$\footnote{Notice that $m_t$ and $m_b$ in composite Higgs models have to be run down from the composite scale, so that their mass intervals  adopted reflect the uncertainties entering such an evolution.}. Also notice that we have contrained the $W^-t\bar b$, $Zt\bar t$ and $Z b\bar b$ couplings to data. Finally,   regarding the latter, the program also checks that the left- and right-handed couplings of the $Z$ boson to the bottom (anti)quark are separately consistent with results of LEP and SLC \cite{Z-Pole}.

As mentioned, in the 4DCHM description, additional fermions  belong to its spectrum, $t'$s and $b'$s,  with SM-like charges.  As these states are heavy quarks, they can in principle be produced in hadron-hadron collisions. The most stringent limits on their mass come at present from the LHC.  To account for the latter, an analysis of the compatibility of the 4DCHM with LHC direct measurements has been performed. The pair production cross section $\sigma(pp\rightarrow t' \bar t'/b'\bar b')$ has been calculated according to the code described in \cite{Cacciari:2011hy}. Clearly, in the 4DCHM, such mass limits would apply to the lightest $t'$ and $b'$ states, i.e., $T_1$ and $B_1$ in Tab.~\ref{table:partspec}. Our limits on $t'$s are based on \cite{CMS:2012ab}, where a search for pair production of $t'$s is performed in CMS with 5 fb$^{-1}$ of  luminosity, where the $t'$s are assumed to decay 100\% into $W^+b$, and on \cite{Chatrchyan:2011ay}, where a search for pair production of $t'$s is performed at CMS with 1.14 fb$^{-1}$ of  luminosity, where the $t'$'s are assumed to decay 100\% into $Zt$. The limits on $b'$s are based on \cite{Chatrchyan:2012yea}, where a search for pair production of $b'$s is performed at CMS  with 4.9 fb$^{-1}$ of luminosity with the $b'$'s that are assumed to decay 100\% into $W^-t$, and on \cite{CMS-PAS-EXO-11-066}, where a search for pair production of $b'$s is performed at CMS with 4.9  fb$^{-1}$ of luminosity with the $b'$'s that are assumed to decay 100\% into $Zb$.  Finally, notice that data considered here come from the 7 TeV run of the LHC. Results for $T_1$ and $B_1$ are shown in Figs.~1
and 2 of Ref.~\cite{Barducci:2013wjc}, respectively\footnote{{More recent results from CMS are given in \cite{Chatrchyan:2012vu} and \cite{Chatrchyan:2012af}, however, they will not change our conclusions.}}.
In practise, from the analysis, limits of about 400 GeV on both $m_{T_1}$ and $m_{B_1}$ can be ascertained.

{{However, one ought to notice that, beside the heavy fermions with ordinary charges, i.e., the $t'$s and $b'$s
in our notation, the composite fermionic spectrum presents also states with exotic charge, as mentioned in the model description. Although these states do not couple directly to the Higgs boson, so that they are inert for our purposes here,
it is important to set bounds on their masses since, in certain region of the parameter space, they can be almost degenerate with the lightest $t'$ or  $b'$.  Since the fermionic spectrum is determined by the  parameters we listed in Sect. II, it is clear that a bound on $\tilde T_1$ (the lightest fermion with  charge $5/3$) mass reflects also on $m_{T_1}$ and $m_{B_1}$.
Regarding  $\tilde T_1$, since in the 4DCHM this particle decays  almost 100\% of the times into $W^+t$, it is possible to apply directly the bound of $650\;{\rm GeV}$ given by \cite{ATLAS:2012hpa}.
Nevertheless, there are regions of the fermion parameter space where the $\tilde T_1$ is not the lightest heavy fermion\footnote{{{In contrast, we have to say that, at the moment, no limits for the charge $-4/3$ fermions are given by the ATLAS and CMS collaborations. They will of course further cut on the low mass values for $T_1$ and $B_1$.}}}}.
This means that the aforementioned values of $m_{T_1}$ and $m_{B_1}$ around 400 GeV remain valid. 

The additional gauge bosons of the 4DCHM, the $W'$s and $Z'$s,  are taken with masses and couplings compliant with experimental limits from both EW precision measurements and direct searches \cite{Barducci:2012kk}.


\section{Results}
\label{Sec:Results}
\noindent

In this section we  compare the yield of the surviving points with the LHC data. To this end,
a useful procedure to adopt is to define the so-called $R$ (sometimes called $\mu$) parameters, i.e., the observed signal (in terms of counted events) in a specific channel divided by the SM expectation:
\begin{eqnarray}\label{R}
R_{YY}=\frac{\sigma(pp\to HX)|_{\rm 4DCHM}\times {\rm BR}(H\to YY)|_{\rm 4DCHM}}{\sigma(pp\to HX)|_{\rm SM}\times {\rm BR}(H\to YY)|_{\rm SM}},
\end{eqnarray}
where $YY$ refers here to any possible Higgs decay channel and in our study we consider $YY$ =
$\gamma\gamma$, $b\bar b$,  $WW$ and $ZZ$. The particles (if any) produced in association with the Higgs boson are here denoted by $X$\footnote{In reality, one should notice that eq.~(\ref{R}) is the limiting case in which sensitivity to the $YY$ decay channel is through only one of the production processes \cite{Spira:1995rr,Kunszt:1996yp}. One should more accurately sum over all of the latter. However, given present data samples and for our purposes, such an approximation is accurate enough.}. For the latest experimental results on such quantities, wherein the label 4DCHM is meant to signify actual experimental data, see Tab.~\ref{tab:R}. 

The relevant (for current LHC data) hadro-production processes at partonic level are
(here $V=W,Z$)
\begin{eqnarray}\label{prod}
gg&\to& H
\quad\quad ({\rm gluon-gluon~fusion}),\\ \nonumber
q\bar q&\to& q\bar q H
\quad\quad ({\rm vector~boson~fusion}),\\ \nonumber
q\bar q(') &\to& V H
\quad\quad ({\rm Higgs-strahlung}).
\end{eqnarray} 
\begin{table}[!t]
\begin{center}
\begin{tabular}{|l|l|l|}
	\hline
	 & ~~ATLAS & ~~~~CMS\\
	\hline
	$R_{\gamma\gamma}$ & $\phantom{-} 1.8 \pm 0.4\phantom{-}$ & $\phantom{-}1.564_{-0.419}^{+0.460}\phantom{-}$ \\
	$R_{ZZ}$           & $\phantom{-} 1.0 \pm 0.4\phantom{-}$ & $\phantom{-}0.807_{-0.280}^{+0.349}\phantom{-}$ \\
	$R_{WW}$           & $\phantom{-} 1.5 \pm 0.6\phantom{-}$ & $\phantom{-}0.699_{-0.232}^{+0.245}\phantom{-}$ \\
	$R_{bb}$		 & $- 0.4 \pm 1.0\phantom{-}$ & $\phantom{-}1.075_{-0.566}^{+0.593}\phantom{-}$ \\
	\hline
\end{tabular}
\end{center}
\caption{LHC measurements of some  $R$ parameters  from  ATLAS \cite{ATLAS-CONF-2012-170} and CMS~\cite{CMS-PAS-HIG-12-045} data. (The CMS numerical values can actually be found in \cite{CMStwiki}.) \label{tab:R}}
\end{table}
For the purpose of our analysis, it is convenient to re-write eq.~(\ref{R}) as follows
\begin{eqnarray}\label{newR}
R_{YY}^{Y'Y'}=\frac{\Gamma(H\to Y'Y')|_{\rm 4DCHM}\times \Gamma(H\to YY)|_{\rm 4DCHM}}
                                 {\Gamma(H\to Y'Y')|_{\rm SM       }\times \Gamma(H\to YY)|_{\rm SM       }}
                         \frac{\Gamma_{\rm tot}(H)|_{\rm SM       }}
                                {\Gamma_{\rm tot}(H)|_{\rm 4DCHM}},
\end{eqnarray}
where $Y'Y'$ {denotes incoming particles participating the Higgs boson production, e.g., $gg$ for the first process  and $VV$ for the other processes in eq.~(\ref{prod}), while $YY$ indicates particles into which the Higgs boson decays}\footnote{Notice that the coupling between the Higgs boson and $W$ or $Z$ intervening in the last two production channels in (\ref{prod}) is the same. However, also notice that, in the 4DCHM, the two couplings $HWW$ and $HZZ$ do not rescale in the same way with respect to the SM ones, in particular for the parameter space investigated here, though the differences will be shown to be small. Hereafter, we will adopt the generic label $V$ to signify either a $W$ or a $Z$.}.

For $YY=\gamma\gamma$, $WW$, $ZZ$ we take the dominant production process to be gluon-gluon fusion (i.e., $Y'Y'=gg$) while for $YY=b\bar{b}$ we assume that Higgs-strahlung dominates (i.e., $Y'Y'=VV$, with the appropriate superposition of $WW$ and $ZZ$).
In other words, we trade a cross section for a width (so-to-say) and this is possible, as we will be carrying out our analysis at lowest order without the presence of radiative corrections due to either Quantum Chromo-Dynamics (QCD) or EW interactions. In fact, following Ref.~\cite{LHC-HXSWG:2012nn}, we can cast eq.~(\ref{newR}) also in the following form:
\begin{eqnarray}\label{R_vs_kappas}
R_{YY}^{Y'Y'}=\frac{\kappa^2_{Y'} \kappa^2_Y}{\kappa^2_H},
\end{eqnarray}
wherein (recall that $VV=WW$ or $ZZ$ and notice that $Y,Y'=b/\tau/g/\gamma/V$)
\begin{eqnarray}\label{kappas}
\kappa^2_{b/\tau/g/\gamma/V}=\frac{\Gamma(H\to b\bar b/\tau^+\tau^-/gg/\gamma\gamma/VV)|_{\rm 4DCHM}}
                                                                  {\Gamma(H\to b\bar b/\tau^+\tau^-/ gg/\gamma\gamma/VV)|_{\rm SM       }},
\end{eqnarray}
\begin{eqnarray}\label{kH}
\kappa^2_H=\frac{\Gamma_{\rm tot}(H)|_{\rm 4DCHM}}
                              {\Gamma_{\rm tot}(H)|_{\rm SM       }}.
\end{eqnarray}

The LHC experiments then perform fits to the $\kappa_i$ coefficients in order to test generic Beyond the SM (BSM) assumptions (for which one or more of the the $\kappa_i$s can differ from 1). However, they generally fix $\kappa_H^2=1$, thereby  assuming that the Higgs width does not change \cite{ATLAS-CONF-2012-170,CMS-PAS-HIG-12-045}\footnote{In fact, the only departure from this  they allow for is to take $\kappa_H^2>1$, corresponding to a value of the Higgs total width in the BSM hypothesis larger than in the SM case, thus accounting for, e.g., invisible Higgs decays that are not captured by standard searches.}. This is a rather restrictive condition since most BSM models predict $\kappa_H^2\ne1$, as the Higgs boson under consideration can mix, in such BSM scenarios, directly with other Higgs boson states or, else, the particles to which it couples can in turn mix. Such effects, whichever their nature, would induce the condition $\kappa_H<1$, as it is the case in the 4DCHM. In fact, we will show later on that  many of the 4DCHM effects enter through such a modification of the Higgs total width.}

\begin{figure}[!t]
\centering
\includegraphics[width=0.40\linewidth]{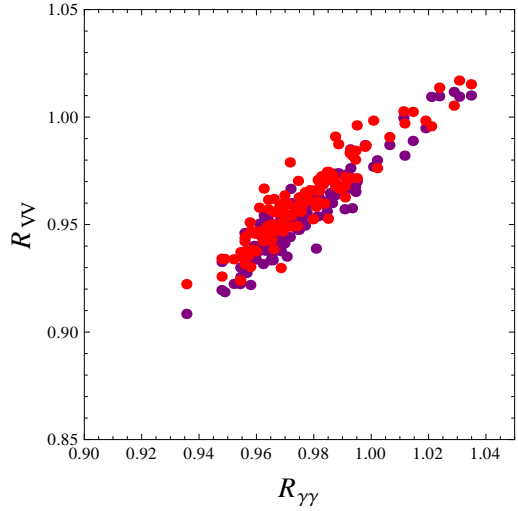}
\caption{Correlation between $R_{\gamma\gamma}$ and $R_{VV}$, with $VV=WW$ (red) and $ZZ$ (purple), from eq.~(\ref{newR}) in the 4DCHM for the benchmark point $f=1$ TeV and $g_*=2$. All points generated here are compliant with direct searches for $t'$s, $b'$s and exotic states with charge 5/3.}
\label{fig:R}
\end{figure}

In order to illustrate the 4DCHM phenomenology, we adopt as reference point the combination $f=1$ TeV and $g_*=2$. However, the salient features extracted for this case may equally be referred to the other benchmark points
to be considered, i.e., those defined in Ref.~\cite{Barducci:2012kk}. Since the errors in Tab.~\ref{tab:R} on $b\bar b$ are very large, Fig.~\ref{fig:R} shows the correlation between the  event rate ratios of eq.~(\ref{newR}) only for the $\gamma\gamma$ and $VV$ channels\footnote{However, the $b\bar b$ case will be taken into consideration later on when making fits to data. In contrast, the $\tau^+\tau^-$ case, being even worse in the above respect, is ignored throughout.}. Furthermore, as intimated, since most of the sensitivity of the $\gamma\gamma$ and $VV$ data is to the gluon-gluon fusion production mode, which is in fact the dominant one in the 4DCHM for the parameter space tested here, like for the SM, we will neglect the effects of all others in our predictions (so that we can conveniently drop the superscripts $gg$ and $VV$ for the time being), except for the $b\bar b$ decay channel (accessible via Higgs-strahlung), that we will consider later on. From the plot in Fig.~\ref{fig:R} it is clear that   there is a noticeable tendency of the model to prefer $R_{\gamma\gamma}$ and $R_{VV}$ values smaller than 1  (the majority of points satisfy this condition, $R_{\gamma\gamma}$ being somewhat larger than $R_{VV}$), with $WW$ showing a slightly stronger propensity than  $R_{ZZ}$ in this direction. The two plotted quantities also appear to be strongly correlated, thereby hinting at a possible common origin for the 4DCHM event rate behaviour relative to the SM predictions. {Moreover, it is also worth mentioning here that the rates for $R_{VV}$ in both the 4DCHM and the SM are computed for the gauge boson decay patterns  which ATLAS and CMS used in reporting the results in Tab.~\ref{tab:R}. These signatures include electrons and muons in all possible combinations entering generic `two-lepton plus missing transverse energy' and `four-lepton' signatures emerging from $WW$ and $ZZ$ pairs, respectively}\footnote{This clarification is of relevance for the case of the 4DCHM, in which the $W$ and $Z$ decay rates change relatively to the SM, unlike the case of other popular BSM models.}. In addition, we have checked that the contribution of $W'$ and $Z'$ states, two of the former and three of the latter, to the yield of these final states, in both mixed 4DCHM/SM and pure 4DCHM channels, is negligible, owing to their large masses as compared to the SM gauge states $W$ and $Z$, despite their large couplings. 

It is now useful to unfold the results in Fig.~\ref{fig:R} in terms of the three $\kappa_i^2$ entering eq.~(\ref{R_vs_kappas}), as each of these can be an independent source of variation in the 4DCHM with respect to the SM. In particular, we map such results in terms of the masses of the lightest $t'$ and $b'$ quarks, i.e., $T_1$ and $B_1$, as these are the 4DCHM quantities to which the event rate ratios are most sensitive. 

\begin{figure}[!t]
\centering
\subfloat[]{
\includegraphics[width=0.45\linewidth]{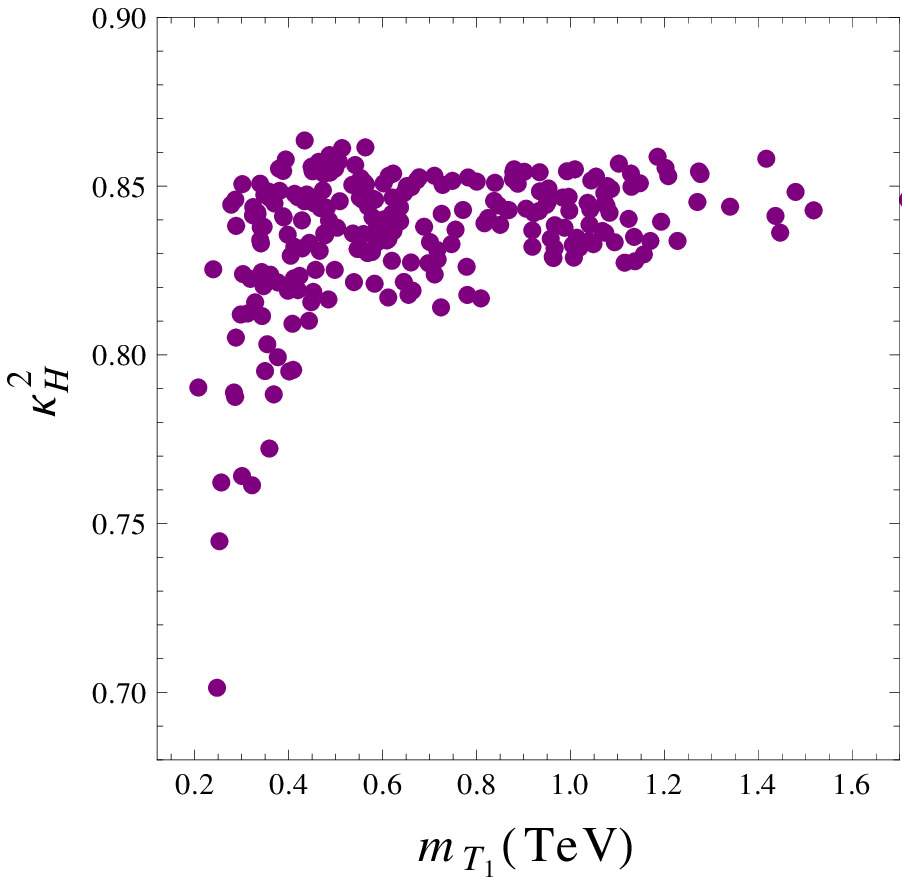}}
\subfloat[]{
\includegraphics[width=0.45\linewidth]{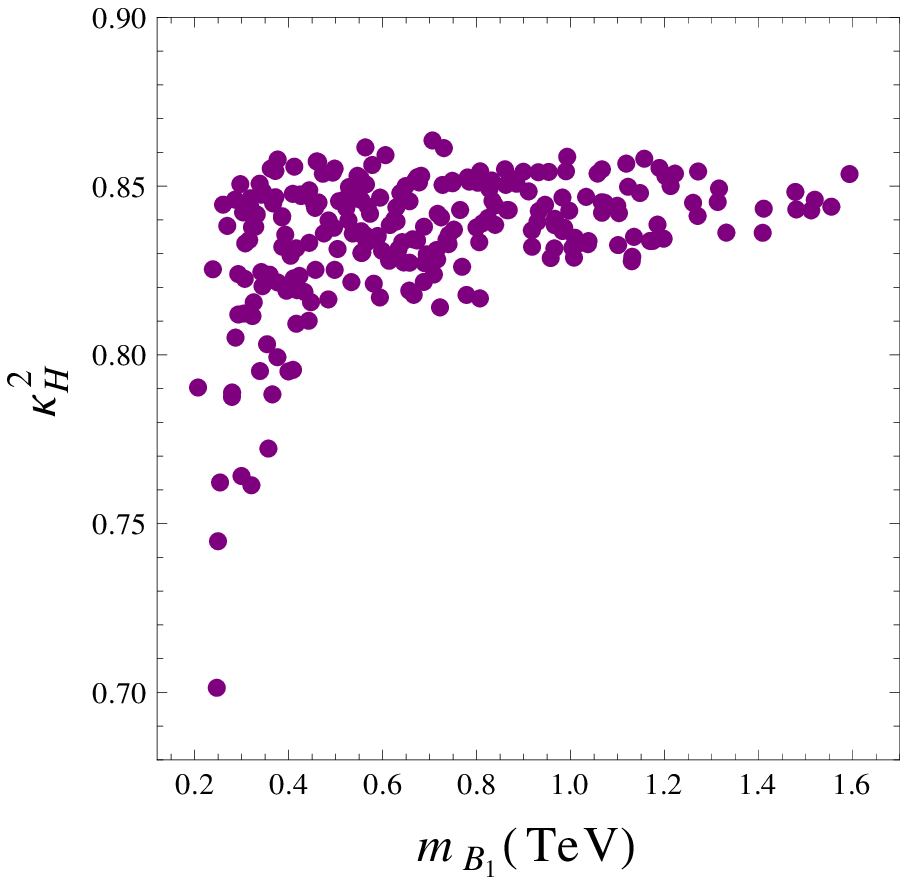}}
\caption{The distributions of $\kappa_H$ values entering eq.~(\ref{R_vs_kappas}) as a function of (a) $m_{T_1}$ and (b) $m_{B_1}$ in the 4DCHM for the benchmark point $f=1$ TeV and $g_*=2$. 
\label{fig:kH}}
\end{figure}

We start with $\kappa_H^2$. This is shown in Fig.~\ref{fig:kH}. Herein, we keep all generated 4DCHM points, including those failing the constraints from direct searches for $t'$, $b'$ states {{or the exotic 5/3 charged fermion}}.  
This is done for the purpose of illustrating the aforementioned sensitivity of the 4DCHM predictions upon the heavy top and bottom masses. In fact, should have smaller $m_{T_1}$ and $m_{B_1}$ been allowed, effects onto the ratio of total widths would have been extremely large, up to $-30\%$ or so. However, even with the aforementioned limits enforced, the 4DCHM effects induced by $t'$ and $b'$ states onto the SM remain substantial, of order $-15\%$ to $-20\%$. Hence, bearing in mind that the contribution of the $H\to gg$, $\gamma\gamma$ and $Z\gamma$ partial widths (those where such $t'$ and $b'$ states enter at lowest order) to the total one are negligible, one has to conclude that these corrections are induced by mixing effects. Furthermore, as $\Gamma_{\rm tot}(H)|_{\rm 4DCHM} \approx\Gamma(H\to b\bar b)|_{\rm 4DCHM}$ (just like in the SM), it is also clear that these are mainly due to $b'$-$b$ mixing affecting the $Hb\bar b$ coupling. Therefore, the result that such 4DCHM effects are negative {is not} surprising. Overall, the reduction of the total Higgs width in the 4DCHM with respect to the SM  {induces}  an increase of all $R$ values in eq.~(\ref{newR}) except, of course, $R_{b\bar b}$.

\begin{figure}[!t]
\centering
\subfloat[]{
\includegraphics[width=0.45\linewidth]{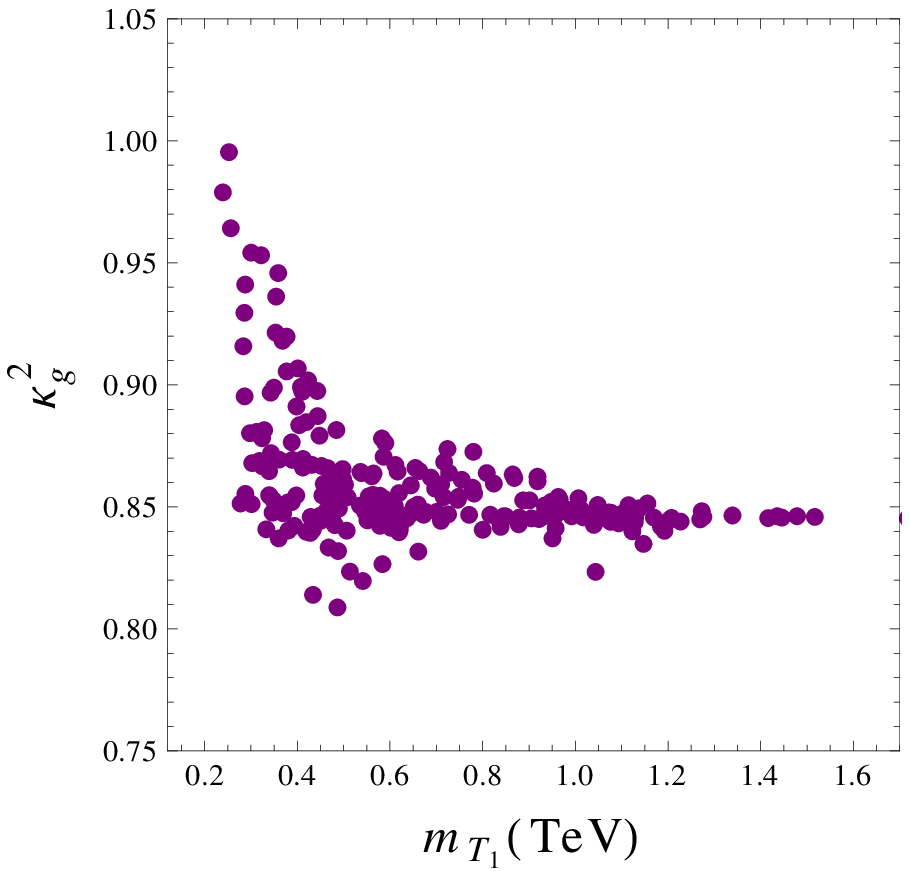}}
\subfloat[]{
\includegraphics[width=0.45\linewidth]{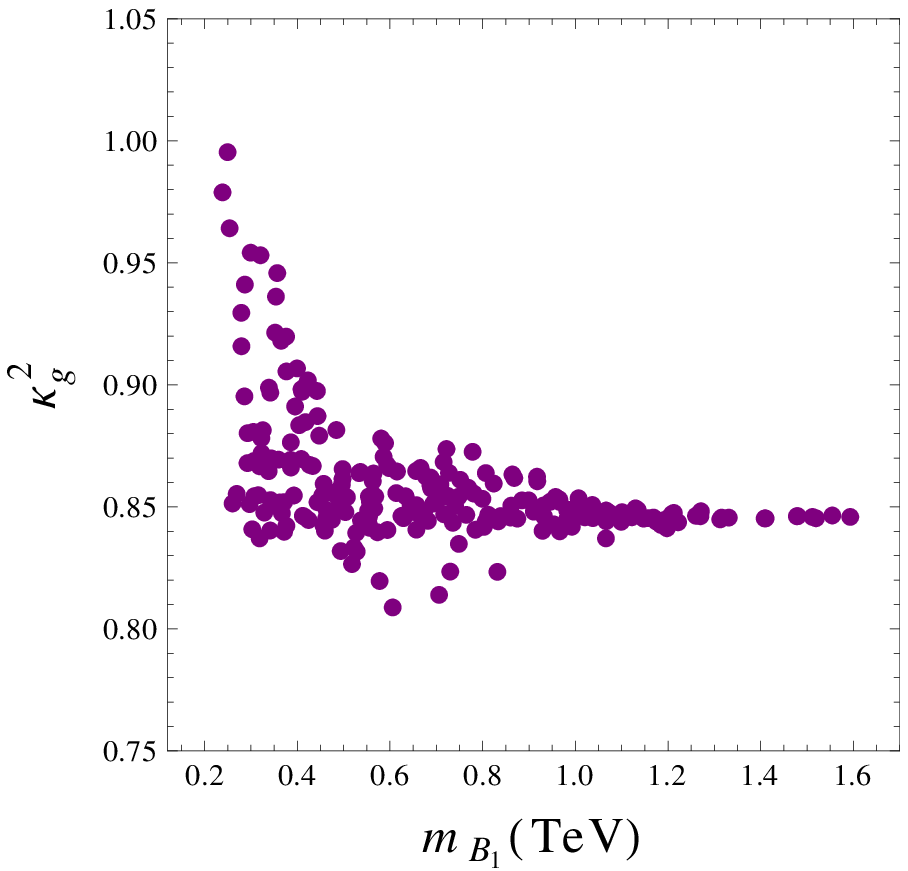}}
\caption{The distributions of $\kappa_g$ values entering eq. (\ref{R_vs_kappas}) as a function of (a) $m_{T_1}$ and (b) $m_{B_1}$ in the 4DCHM for the benchmark point $f=1$ TeV and $g_*=2$. 
\label{fig:kg}}
\end{figure}

\begin{figure}[!t]
\centering
\subfloat[]{
\includegraphics[width=0.45\linewidth]{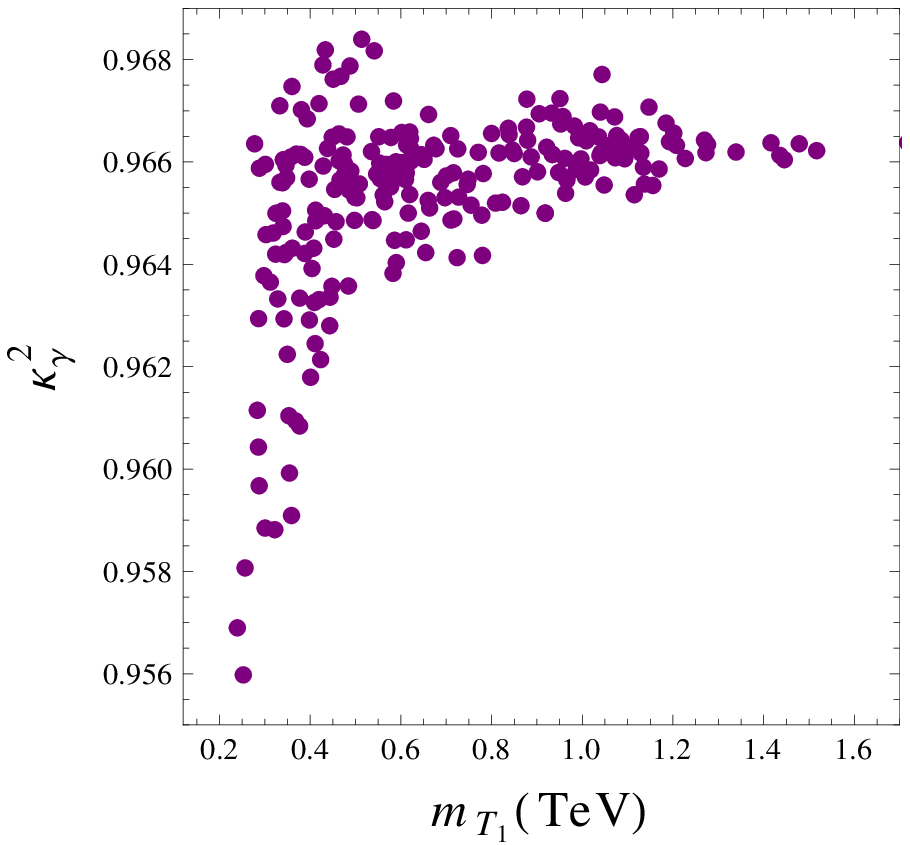}}
\subfloat[]{
\includegraphics[width=0.45\linewidth]{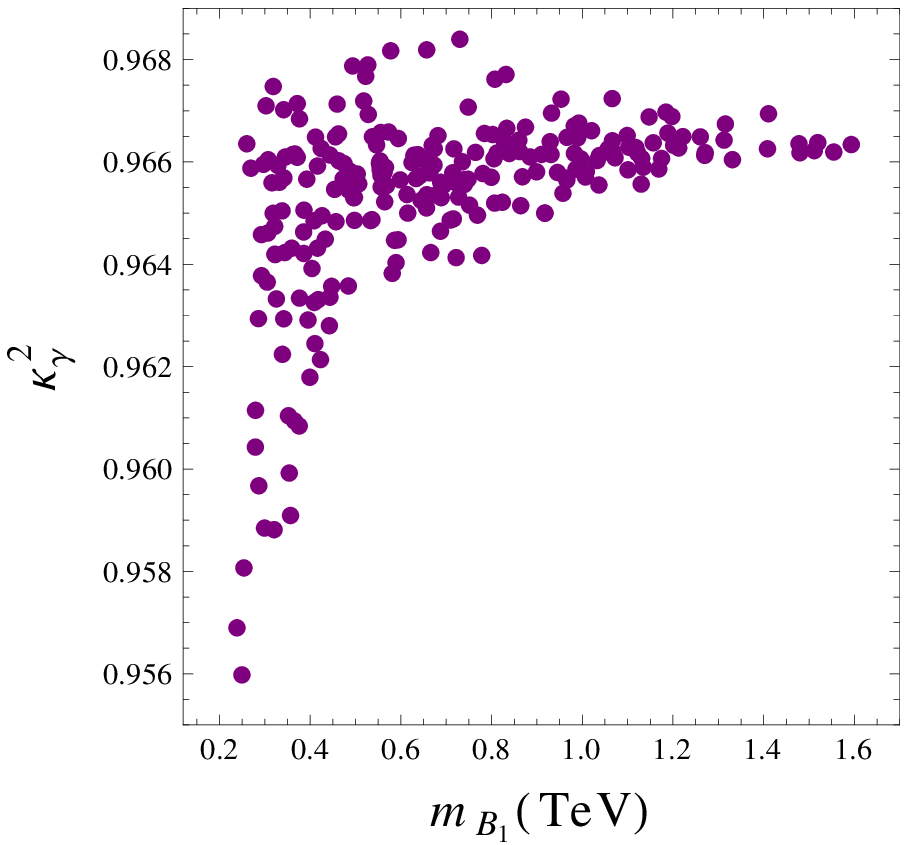}}
\caption{The distributions of $\kappa_\gamma$ values entering eq.~(\ref{R_vs_kappas}) as a function of (a) $m_{T_1}$ and (b) $m_{B_1}$ in the 4DCHM for the benchmark point $f=1$ TeV and $g_*=2$. 
\label{fig:kA}}
\end{figure}

Since we are interested in probing the 4DCHM hypothesis as an explanation of the LHC data used for the Higgs search and since the largest anomaly with respect to the SM is seen in the di-photon channel (recall Tab.~\ref{tab:R}), we next study $\kappa_g^2$ and $\kappa_\gamma^2$, also entering eq.~(\ref{R_vs_kappas})\footnote{The case of $\kappa_V^2$ is relevant too, as also the $HVV$ couplings in the 4DCHM change from their SM values (and, as mentioned, differently for $WW$ and $ZZ$). However, here, the dynamics occur at tree-level, so the effects are trivial, as they can be easily accounted for by replacing the $HVV$ couplings of the SM with those of the 4DCHM. Needless to say, also in this case the differences between SM and 4DCHM are negative and due to mixing, which is non-negligible.}. We show these two quantities in Figs.~\ref{fig:kg}--\ref{fig:kA}, respectively. In both cases, we see a reduction of the partial widths in the 4DCHM relative to the SM. Again, we trace this to mixing effects, this time between $t$ and $t'$ states. Both at production and decay level, in fact, $t$ contributions are larger than the $b$ ones, both in the 4DCHM and SM. {Again, they induce negative corrections, typically $-10\%$ for $\kappa_{g}$ and $-6\%$ for $\kappa_{\gamma}$.} Furthermore, that the former are larger than the latter is due to the fact that the $t$ loop is the leading one in the production graph whereas it is subleading (smaller than the $W$ contribution) in the decay diagram. Incidentally, unlike the case of the total width, for these two partial widths, if lighter $T_1$ and $B_1$ masses were allowed (they are strongly correlated), genuine 4DCHM effects onto the SM would have been different in the two channels, typically inducing larger(smaller) rates at production(decay) level. In this  dynamics we recognise the effects of the $t'$ and $b'$ loops in the two triangle amplitudes (as opposed to those induced by mixing in the couplings). In fact, the lighter the $t'$ and $b'$ masses, the bigger their loop contributions\footnote{Also recall that the $Ht'\bar t'$ and $Hb'\bar b'$ couplings are not of Yukawa type, that is, they do not scale linearly with the $t'$ and $b'$ masses.}. {So that, at both production and decay level, the net effect from $t'$ and $b'$ loops turn out to have the same sign as the $t$ one (recall that the $b$ ones are negligible in both models) for the case of a light $T_1$ or $B_1$ (below 500 GeV). Hence, in production they interfere constructively with the leading $t$ contribution, which in turn means that they interfere destructively in decay with the leading $W$ contribution (which has a sign opposite to the $t$ term). In case of a heavier $T_1$ and $B_1$, the sign of the overall contribution of $t'$ and $b'$ quarks can vary with respect to the top quark one but the combined contribution of all heavy quarks  is quite small. In fact, we have verified that the asymptotic values, i.e., for large $m_{T_1}$ and $m_{B_1}$, attained by $\kappa_g^2$ and $\kappa_\gamma^2$ in Figs.~\ref{fig:kg} and \ref{fig:kA}, respectively, coincide with those obtained in the aforementioned literature by adopting the described decoupling approximation of the heavy fermionic states. Conversely, it should be noted that the asymptotic results can differ significantly from those obtained for small $T_1$ and $B_1$ masses, particularly for $\kappa_g^2$, up to $7\%$ or so (around 400 GeV). For smaller masses, the effect would be even more prominent.}

To summarise then, we are  in presence of contrasting effects entering eq.~(\ref{R_vs_kappas}). All $\kappa_i^2$ therein tend to diminish, relative to the SM. {However,  the decrease of $\kappa_H^2$ entering the denominator is bigger than the decrease of the $\kappa_Y^2\times\kappa^2_{Y'}$ product in the numerator, so that the net effect could be the increase of  the event rate in comparison to the SM. This dynamics was indeed shown in Fig.~\ref{fig:R} while its details can be seen in Figs.~\ref{fig:kH}--\ref{fig:kA}}.

\begin{figure}[!t]
\centering
\subfloat[]{
\includegraphics[width=0.45\linewidth]{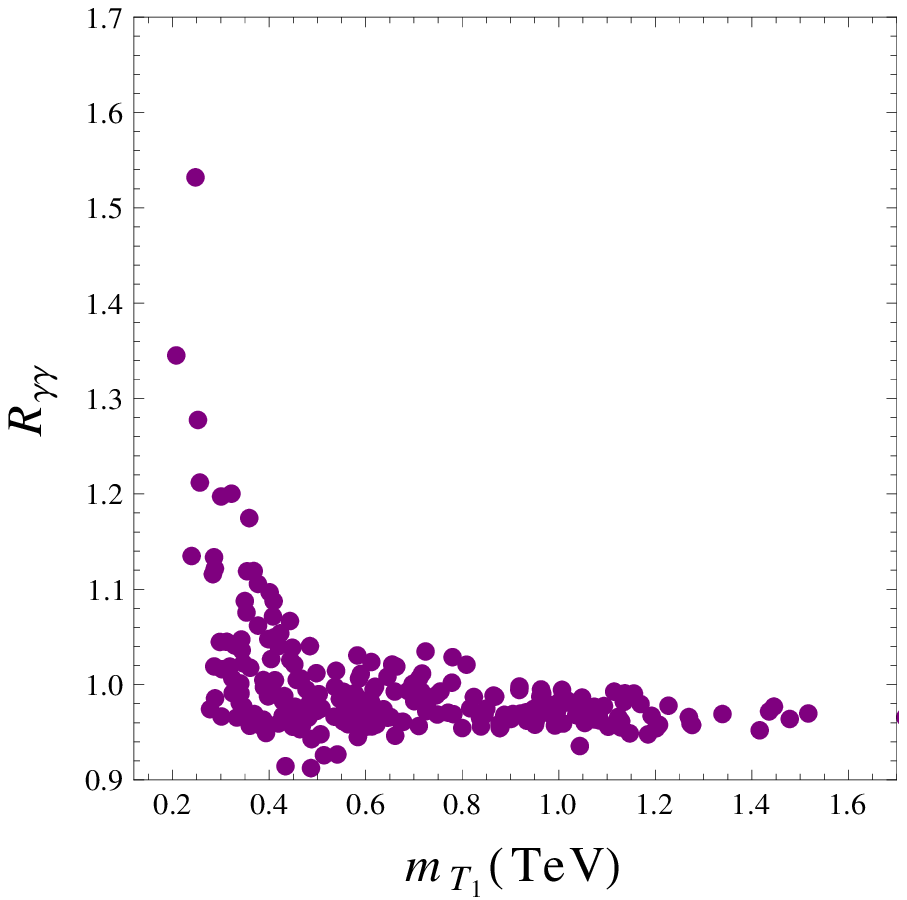}}
\subfloat[]{
\includegraphics[width=0.45\linewidth]{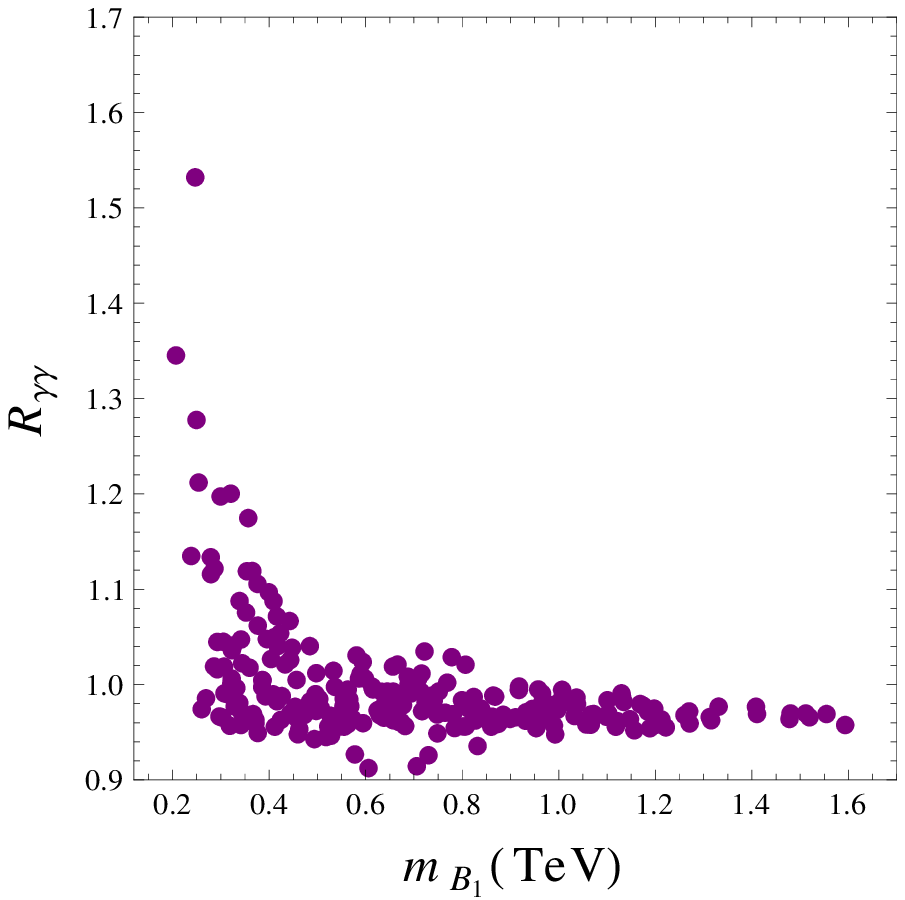}}
\caption{The distributions of $R_{\gamma\gamma}$ values entering eq.~(\ref{newR}) as a function of (a) $m_{T_1}$ and (b) $m_{B_1}$ in the 4DCHM for the benchmark point $f=1$ TeV and $g_*=2$. 
\label{fig:RA}}
\end{figure}

{In Fig.~\ref{fig:RA} we investigate these effects further for $R_{\gamma\gamma}$, for which (recall) the largest anomaly is seen, plotted as a function of $m_{T_1}$ and $m_{B_1}$. For $m_{T_1}$ and $m_{B_1}$ above 400 GeV, $R_{\gamma\gamma}$ values can reach 1.1. However, again, should heavy quark masses be allowed to be below 400 GeV or so, the values for $R_{\gamma\gamma}$ could have been rather large, up to 1.2 (or even more). In fact, quite irrespectively of the actual values attained by $R_{\gamma\gamma}$, the tendency in Fig.~\ref{fig:RA} is clear enough.} There is a consistent `leakage' of points towards $R_{\gamma\gamma}>1$, the more so the lighter $m_{T_1}$ and $m_{B_1}$. The relevance of this result is twofold. On the one hand, this calls for a thorough re-examination from an experimental point of view of the actual limits on the $t'$ and $b'$ states, especially for low masses, certainly affording an accuracy well beyond the one stemming from the rudimentary approach we have adopted in Figs.~1--2
of Ref.~\cite{Barducci:2013wjc}. {On the other hand,  we would like to argue that statements from previous literature, mentioning that accurate predictions can be made in the infinite $t'$ and $b'$  mass limit \cite{Azatov:2011qy,Azatov:2012rd,Gillioz:2012se,Azatov:2012ga}, may not be applicable to our concrete realisation of the 4DCHM.}

However, for the time being, we take the limits on $m_{T_1}$ and $m_{B_1}$ as we obtained them at face value and collect all the results produced, including those for the other $f$ and $g_*$ benchmarks, and compare them to the experimental results of ATLAS~\cite{ATLAS-CONF-2012-170} and CMS~\cite{CMS-PAS-HIG-12-045}  collected in Tab.~\ref{tab:R}.

\begin{figure}[!t]
\centering
\subfloat[]{
\includegraphics[width=0.35\linewidth]{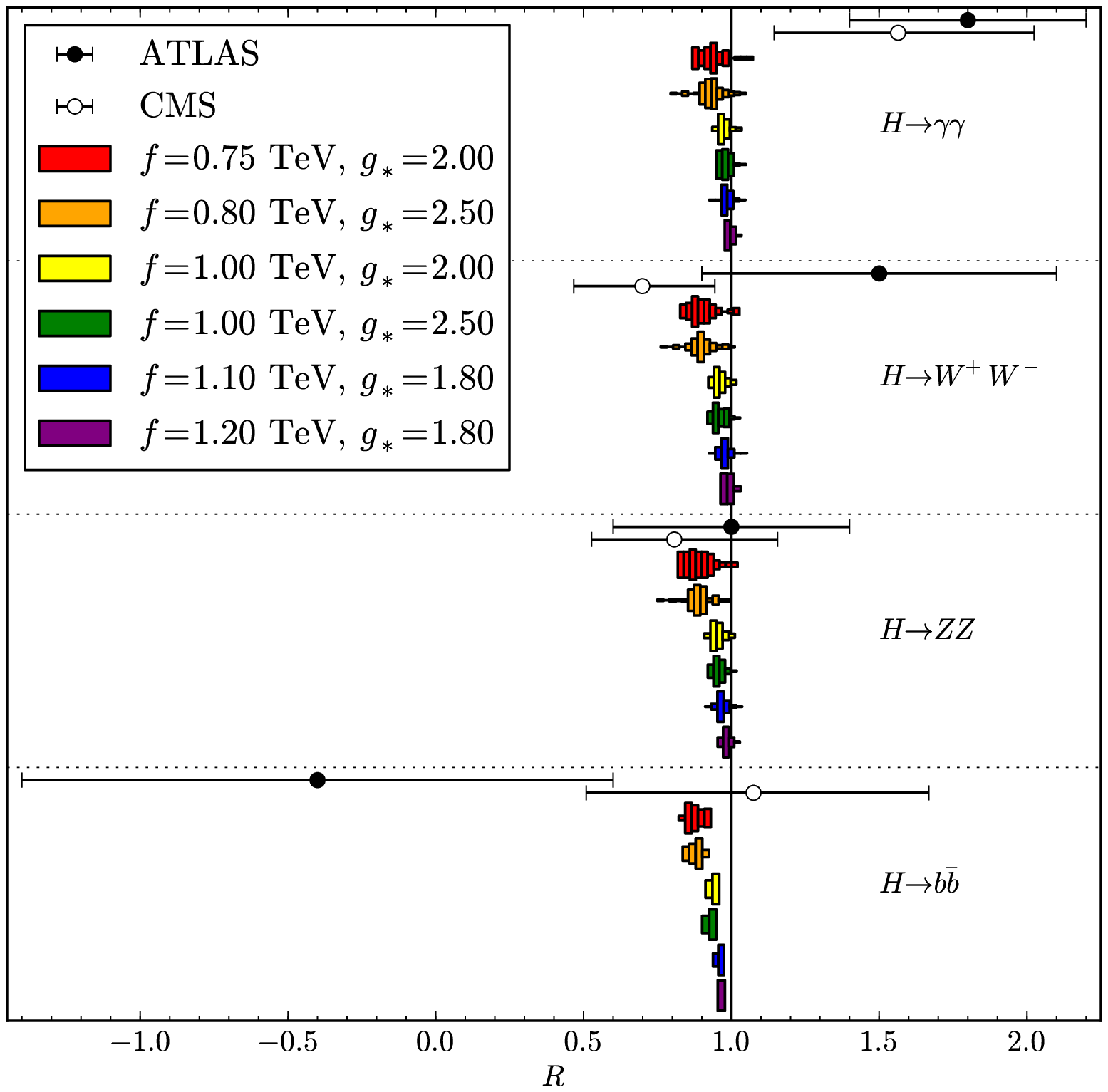}}
\subfloat[]{
\includegraphics[width=0.45\linewidth]{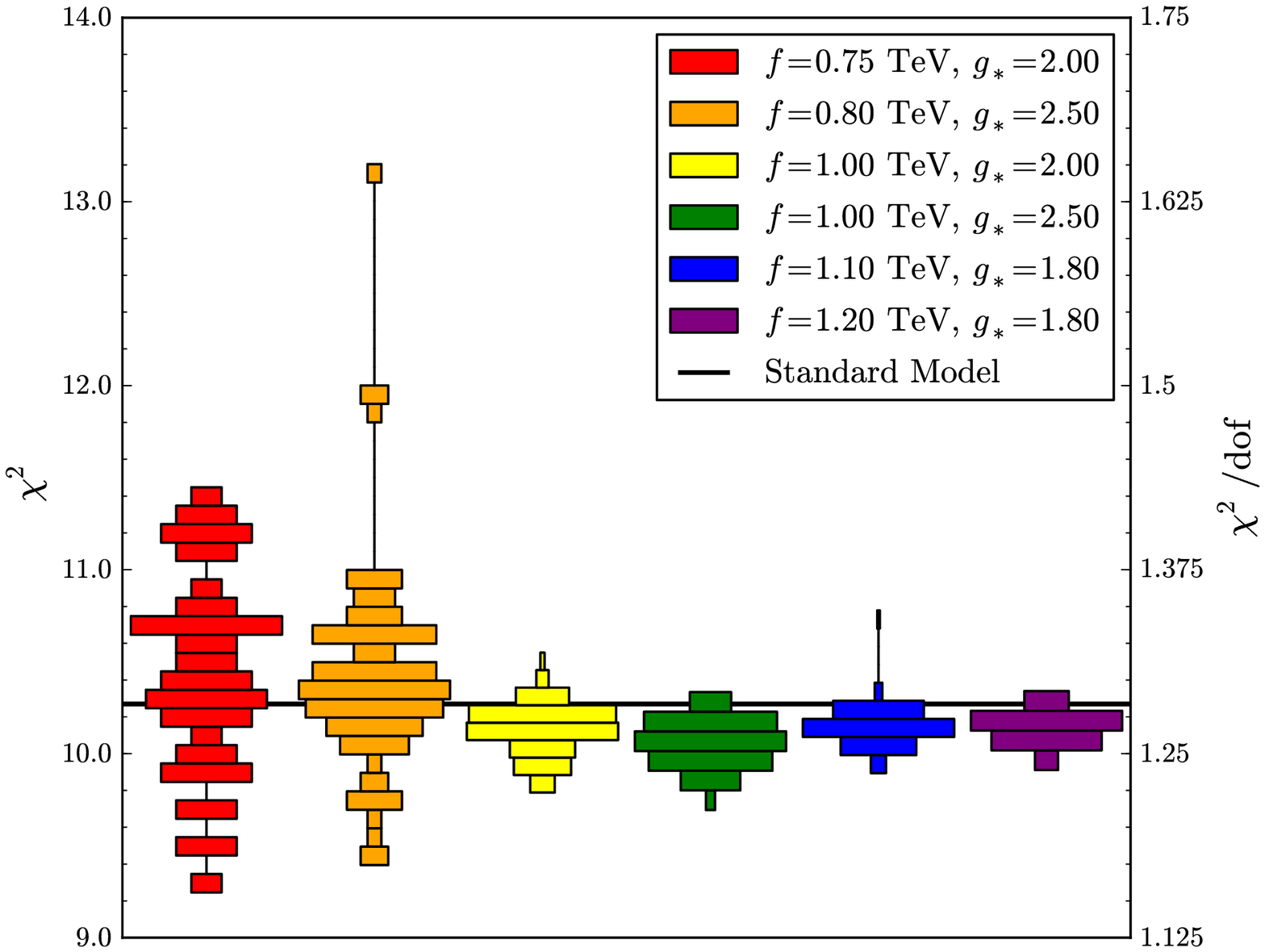}}
\caption{(a) Comparison of the $R$'s from eq.~(\ref{newR}) with the measured experimental values by ATLAS~\cite{ATLAS-CONF-2012-170} and CMS~\cite{CMS-PAS-HIG-12-045} (see Tab.~\ref{tab:R})  in the 4DCHM for all benchmark points in $f$ and $g_*$.  (b) The $\chi^2$ fit (as described in the text) in the 4DCHM for all benchmark points in $f$ and $g_*$. All points generated here are compliant with direct searches for $t'$s, $b'$s and exotic states with charge 5/3.}
\label{figs:Matt}
\end{figure}

For each $(f,g_*)$ benchmark we scan over the other free parameters and remove points that do not survive the $t^\prime$, $b^\prime$ and charge 5/3 quark direct search constraints. For the remaining points we calculate $R_{YY}$ for $YY=\gamma\gamma$, $WW$, $ZZ$, $b\bar{b}$. The results are shown in Fig.~\ref{figs:Matt}(a) as a series of normalised histograms in order to demonstrate the number of points in the scan taking particular values of $R_{YY}$ and the full range of values obtained is shown. The experimental measurements for $R_{YY}$ are shown by black and white points with 68\% Confidence Level (CL) error bars. As the scale $f$ is increased, the values of $R_{YY}$ become more sparse. This is because parameters in the model become more tightly constrained as this scale grows larger.

{In order to have a clear picture on  how the 4DCHM fares against LHC data, particularly in relation to the SM, in a  quantitative way, we calculate the $\chi^2$ goodness of our fit for the ATLAS~\cite{ATLAS-CONF-2012-170} and CMS~\cite{CMS-PAS-HIG-12-045} data from Tab.~\ref{tab:R}}. We assume that all the channels and experiments are independent and simply sum them in the $\chi^2$ function, giving us eight degrees of freedom (dof). The value of $\chi^2$ for each parameter scan point (surviving the experimental constraints discussed above) is shown in Fig.~\ref{figs:Matt}(b) using normalised histograms. The value of our $\chi^2$ function for the SM (i.e., $R_{YY} = 1$) is also plotted as a horizontal black line and the figure makes it clear that the 4DCHM represents a better fit to the data than the SM does for most of the benchmarks considered.


\section{Conclusions}
\label{Sec:Conclusions}
\noindent

{In summary, we have shown that the 4DCHM could provide an alternative (at times even better) explanation  than the SM of  the current LHC data pointing to the discovery of a neutral Higgs boson with mass around  125 GeV. 

After systematically scanning the parameter space of the 4DCHM and illustrating its phenomenology for several benchmark points, we have shown that a moderate enhancement  in  the $H\to \gamma\gamma$ channel with respect to the SM predictions is a possible feature of this model and can be as large as about 1.1, somewhat below the central values of the experimental measurements.

However, we have also found that this enhancement could  potentially be larger, realistically up to 1.3, due to the contribution from the  heavy $t'$ and $b'$ fermions of the 4DCHM  with mass just below 400 GeV,  i.e.,  precisely when entering mass regions apparently excluded but for which there are no data from direct searches, only simple extrapolations that we attempted here. So, a thorough re-assessment from the experimental side is required in this respect\footnote{After the release of \cite{Barducci:2013wjc}, we have become aware that some experimental work in this direction has started
\cite{ATLAS-CONF-2013-018}, which, however, does not change the main conclusions of this paper.}.

The main source of the enhancement of the $H\to \gamma\gamma$ channel is in the reduction of the $H\to b\bar{b}$ partial width due to $b$-$b'$ mixing effects which in turn leads to the reduction of the total Higgs boson width and the enhancement of  all decay channels, including the di-photon one. Competing effects emerge though from the (effective) $Hgg$ coupling becoming simultaneously smaller.

Finally, a relevant by-product of our analysis has been to show that several approximations adopted in literature 
in scenarios similar to the 4DCHM, which essentially make predictions in the limit in which the masses of the heavy fermions (and possibly heavy gauge bosons) are infinitely heavy, cannot generally be accurate over the entire parameter space of the corresponding model. 
 

\begin{acknowledgments}
SM is grateful to the workshop organisation for financial support and for a stimulating and entertaining meeting.
DB, AB and SM are financed in part through the NExT Institute. The work of GMP has been supported by the German Research Foundation DFG through Grant No.\ STO876/2-1 and by BMBF Grant No.\ 05H09ODE.
\end{acknowledgments}

\bigskip 
\bibliography{HPNP2013_Moretti}

\end{document}